\documentclass[sigconf,nonacm]{acmart}
\usepackage{tabularx}
\usepackage[utf8]{inputenc}
\usepackage{dsfont}

\date{November 2021}

\usepackage{natbib}
\usepackage{graphicx}

\acmDOI{}

\acmConference[]{}

\begin{document}

\title{Community-Detection via Hashtag-Graphs for Semi-Supervised NMF Topic Models}

\author{Mattias Luber}
\orcid{0000-0002-3775-0804}

\author{Anton Thielmann}
\email{}
\orcid{0000-0002-6768-8992}

\author{Christoph Weisser}
\email{}
\orcid{0000-0003-0616-1027}

\author{Benjamin Säfken}
\email{}
\orcid{0000-0003-4702-3333}

\renewcommand{\shortauthors}{}

\begin{abstract}
    Extracting topics from large collections of unstructured text-documents has become a central task in current NLP applications and algorithms like NMF, LDA as well as their generalizations are the well-established current state of the art. However, especially when it comes to short text documents like Tweets, these approaches often lead to unsatisfying results due to the sparsity of the document-feature matrices.
    Even though, several approaches have been proposed to overcome this sparsity by taking additional information into account, these are merely focused on the aggregation of similar documents and the estimation of word-co-occurrences. This ultimately completely neglects the fact that a lot of topical-information can be actually retrieved from so-called hashtag-graphs by applying common community detection algorithms. Therefore, this paper outlines a novel approach on how to integrate topic structures of hashtag graphs into the estimation of topic models by connecting graph-based community detection and semi-supervised NMF. 
    By applying this approach on recently streamed Twitter data it will be seen that this procedure actually leads to more intuitive and humanly interpretable topics.
\end{abstract}

%



\maketitle

\section{Introduction}
Topic modelling describes the extraction of hidden semantic structures that occur in large sets of text documents, where these structures represent human-understandable topics. Two algorithms suitable and commonly used for this task are Latent Dirichlet Allocation (LDA) \cite{Blei.2003} and non-negative matrix factorization (NMF) \cite{Cedric.2010, Pedregosa.2011}. However, especially for short text documents, LDA can lead to unsatisfactory results \cite{Maz16} due to the sparsity of the data and the inherent assumption, that documents are merely a mixture of latent topics. \cite{Chen.2019} conducted a comprehensive exploratory study between various LDA and NMF based schemes, which concluded that NMF indeed outperforms LDA for extracting latent topics in short texts. This results from the fact that short documents lead to extremely sparse document-feature matrices and therefore word-word co-occurrences are harder to estimate. 
There are approaches better suited for short text clustering than LDA like the Gibbs Sampler Dirichlet Multinomial Model (GSDMM) and the Gamma Poisson Mixture Model (GPM) \cite{Yin14, Maz20}. These approaches are trying to fix the LDA inherent assumption of having multiple topics per document by using Dirichlet-Mixture-Models instead. Other approaches try to improve the estimation of word-word co-occurrences by e.g. using word embeddings. Many of these can be found in \cite{Chen.2019}, where their performance on short documents is investigated.

None of these approaches, however, include possibly available additional information in their modelling as for example the structural or the author-topic model \cite{roberts2014structural, rosen2012}. Although, in what is probably the world's largest collection of short-text documents, namely Twitter data, additional valuable information about the latent topics covered in the Tweets is included in the form of hashtags.

In the literature a wide range of approaches have been proposed to overcome the tweet-specific sparsity. The most simplistic one is to aggregate several Tweets into longer pseudo-documents based on common attributes such as hashtags (tweet pooling) \cite{Mehrotra.07282013}. While the available additional information is used in this approach, it is not included in the actual modelling and only a means of creating longer documents that are better suited for the LDA algorithms. Additionally, even though \cite{Mehrotra.07282013} showed that this method can indeed improve the estimations for LDA it raises another problem for large collections of Tweets: Especially for general purpose hashtags which are heavily and constantly used across time, this aggregation can lead to a few extraordinarily large documents, while rare hashtags are still producing short documents. 
An introduction of more advanced methods to use the information carried by hashtags is for example done in \cite{wang2011topic} or \cite{wang2014hashtag}. \cite{wang2011topic} use hashtag graphs for sentiment analysis, by taking the sentiment of neighbouring hashtags into account. \cite{wang2014hashtag} is integrating hashtag co-occurrences into a probabilistic model to improve topic estimates, but besides that they are not actually making use of the actual structure of the graph or of any properties of graphs in general. In contrast, we are applying graph algorithms explicitly suited for capturing the structure in graphs, i.e. community detection in the hashtag graph. We find that the detection of topics works very well with hashtag graphs and provides valuable information about the topics that are actually "hidden" in the data. Hence, we use the topics obtained from the community detection in the hashtag graph as prior information for semi-supervised NMF to improve estimates. Thereby, we propose a novel method of utilizing the stored information carried in the hashtags into document topic extraction by connecting graph-based community detection and semi-supervised NMF. The remainder of the paper is structured as follows: In chapter \ref{chap:nmf} a brief introduction to NMF is given. The construction and analysis of the created hashtag graphs is explained in chapter \ref{chap:hastag}. Following, the integration of both methods into the described approach is derived. In chapter \ref{application}, the method is applied to real-world twitter data. Lastly, a short evaluation of the results and conclusion is given.

\section{Topic modelling with Non-Negative Matrix Factorization}\label{chap:nmf}

Let $\textup{V} =\left \{ v_1,~...~, v_w   \right \}$ be the vocabulary of tokens and  $\textup{D} =\left \{ d_1,~...~, d_n    \right \}$ be the collection of documents. Each document is then assumed to be a sequence of words $d_i = \left [ v_{i1}, ..., v_{iw_i} \right ]$  where $v_{ij} \in V$ hold and where $w_i$ denotes the length of document $d_i$. Based on that the data matrix (sometimes called document-feature-matrix) $X \in \mathbb{R}^{n \times w}$ is constructed, by applying word-count-factorization. 

NMF is often used with a TF-IDF (term frequency/inverse document frequency) reweighting scheme $f(X)$ such that $\widetilde{X} = f(X)$ is used for the topic modelling. This can help to retrieve the meaningful words for each document since it lowers the coefficients for tokens that are either used across too many documents or barely used at all.
Based on that, the algorithms then try to identify hidden topics.  

Generally, NMF is an algorithm where an input matrix is factorized into multiple (most commonly two) matrices that do not contain any negative elements.
More specifically, for the case of two factorized matrices, it tries to approximate a data matrix, $X$, with a product of a coefficient matrix and a component matrix as $X\approx WH$ for a given number of components $k$. For document clustering, the input matrix $X \in \mathbb{R}^{n\times w}$ is hence a word-document matrix with the words stored in the rows and the documents stored in the columns. The matrices $ W \in \mathbb{R}^{n \times k}$ and $H \in \mathbb{R}^{k \times w}$ are derived by minimizing the reconstruction error in respect to e.g. the Frobenius norm with the additional side constraint, that all coefficients have to be non-negative, i.e.

$$ (W,H) = arg ~min \left \| \widetilde{X} - WH \right \|_F ~ with ~ W, H \geq  0.$$

\noindent In the end, the documents are then expected to be linear combinations of the components and due to the non-negativity of the coefficients, the rows of $H$ become interpretable as topics.

\section{Construction and Analysis of Hashtag Graphs}\label{chap:hastag}
In the following, the construction of a hashtag graph is defined, based on which topic information is derived by the use of community detection algorithms.
Tweets have the convenient property that in many cases they already come with some kind of topical information, since the user can actively soft-label them with hashtags. Since users are not restricted in the choice and number of hashtags per Tweet, the hashtags can not be used as labels directly. Still, those tags already reveal valuable information about the topic a certain Tweet tries to address and it will be seen that community detection in hashtag graphs are surprisingly effective to gather this structure.

\subsection{Notation and Definition
}
\begin{figure}[H]
    \centering
    \includegraphics[scale=0.6]{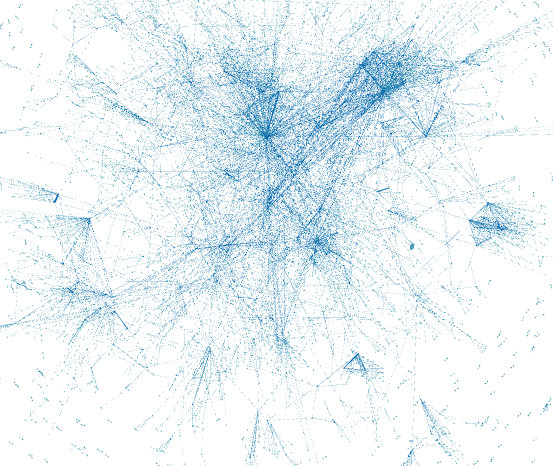}
    \caption{Visualization of the hashtag graph. It can be seen that the graph appears to be structured in well connected areas, which are rather loosely connected to the rest of the graph. Those  dense areas are referred to as communities.}
    \label{fig:full_graph}
\end{figure}
Let $\textup{V} =\left \{ v_1,\ldots, v_w   \right \}$ be the vocabulary of tokens over all Tweets and $\textup{T} =\left \{ t_1,\ldots, t_h   \right \} \subset V$ the set of hashtags. Furthermore, the set of documents is denoted as $\textup{D} =\left \{ d_1,\ldots, d_n    \right \}$ such that $d_i \subset V$, where each Tweet makes up a document. 

\noindent The weighted undirected hashtag-graph can then be defined as a set of edges:

$$\textup{E} = \left \{ (t_i,t_j)\mid \exists d_k\in D: t_i,t_j \in d_k \mbox{ for } t_i,t_j \in T \right \},$$ 

\noindent with a weighting-scheme such that for $e_{ij} = (t_i,t_j) \in E$ the corresponding weight is defined by 

$$\phi_{ij}  = \left | \left \{ d_k\mid t_i,t_j \in d_k,\ d_k \in D\right \} \right |.$$

\noindent Informally speaking, this leads to a graph where each node represents a hashtag and a connection is drawn between two nodes if at least one Tweet exists where those tags co-occur. The weighting represents the number of Tweets where those tags co-occur.

To increase the robustness, a threshold $\tau \in \mathbb{N}$ is introduced such that all edges are excluded whose corresponding weight fall short of this threshold, i.e. 
$$ \textup{ E }_{trunc} = \left \{ e_{ij}\in E \mid \phi_{ij}  \geq \tau \right \}.$$
This is justified by the assumption that meaningful connections  between hashtags will occur more or equal then $\tau$ times and helps to reduce the computational complexity of graph operations.

\subsection{Network Analysis and Community Detection}
In the next step, the hashtag graph is analysed and community detection is performed to extract clusters of hashtags, that often are used in the same context. The clusters are then later used to label the data for the topic modelling.
For the analysis and visualization of the graph the open source platform Gephi \cite{MathieuBastian.2009} and the python library networkx \cite{AricA.Hagberg.2008} are used. Without going into depth it shall be mentioned here that the hashtag graph actually shares a lot of properties with so called small-world networks\cite{telesford2011ubiquity}. The main three shared properties that are relevant for our analysis are:
\begin{itemize}
\item Small-world networks are structured by an unusually high number of community-like sub-graphs. Within those sub-graphs, the nodes are highly connected with each other, but connections to nodes from other sub-graphs are be observed less often.
\item Furthermore, the networks contain an unusually high amount of hubs. Those hubs are nodes which are directly connected to an unusually high number of other nodes.
\item The shortest path between any two nodes is surprisingly short, even for huge networks.
\end{itemize} 

In the context of hashtag graphs this means that the network could be factored into sub-groups of tags which are more likely to occur together in a Tweet and less likely to be used with tags of other groups.

And indeed those properties seem to be pretty accurate by looking at figure \ref{fig:full_graph}. 
\begin{figure}
    \centering
    \includegraphics[scale=0.35]{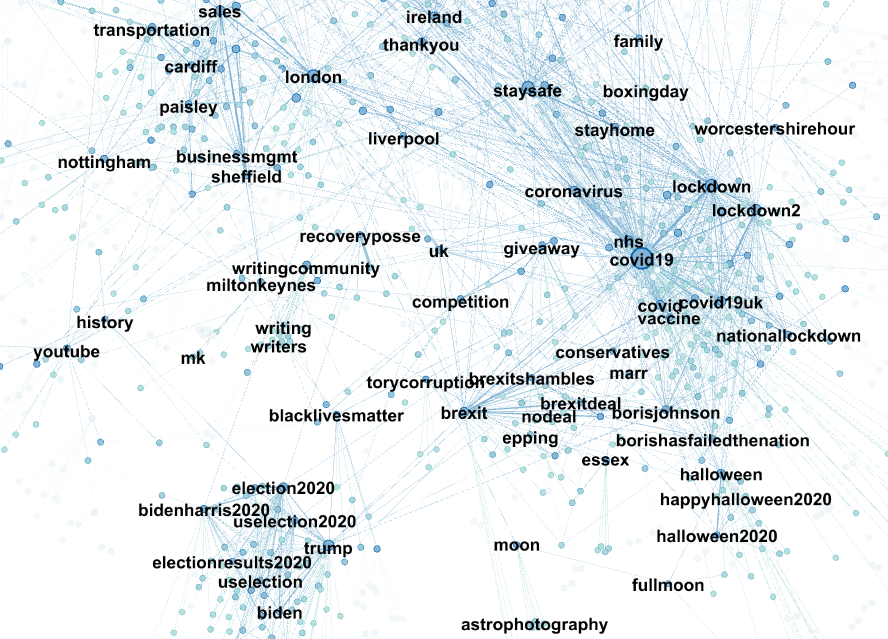}
    \caption{Close up of the hashtag-graph with some nodes assigned with their names. The figure shows that hashtags which are are used in context of the same topic are also ending up close to each other in the graph.}
    \label{fig:graph_labeled}
\end{figure}
By inspecting some of those sub-groups and their labels, it becomes clear that those groups are representing topic-like structures, which is actually not very surprising since hashtags are specifically used to label Tweets. For example, a tweet about the U.S. election would be much more likely to include hashtags about Trump and Biden than about Biden and Halloween.

\begin{figure*}
    \centering
    \includegraphics[scale=0.5]{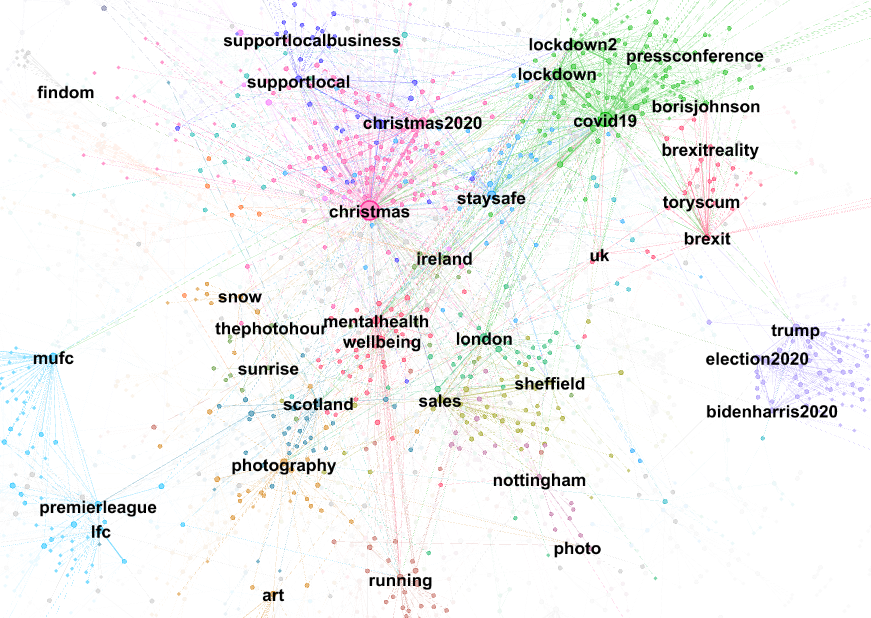}
    \caption{Close up of the hashtag-graph with some nodes assigned with their names and colored based on their determined community. This figure shows that community detection can be used to capture topic structures in hashtag graphs. Hashtag that are showing up in the same community are targeting the same topic in the vast majority of the cases}
    \label{fig:graph_communities}
\end{figure*}

The next step is to retrieve and extract those structures and due to the properties of small-world networks this can easily be done by the Louvain algorithm \cite{VincentDBlondel.2008}. Louvain works by maximizing the modularity score $Q$, i.e.

$$Q = \frac{1}{2m}\sum_{i,j = 1}^{|T|}\left [ (\phi_{ij} -\frac{k_i
k_j}{2m})\cdot \mathds{1}(c_i=c_j) \right ]$$

$$m=\sum_{i,j = 1}^{|T|}\phi_{ij} ,~ k_i = \sum_{j=1}^{|T|} \phi_{ij} ,~ k_j = \sum_{i=1}^{|T|} \phi_{ij} $$

%

\noindent where $c_i,~ c_j$ are the communities of $t_{i}$ and $t_{j}$. $k_i$ and $k_j$ are the weights of the nodes $t_i$, $t_j$ determined by summing up the weights of all related edges.
Informally speaking, $Q$ compares the degree of connection within the communities, with the degree of connection between communities.
$m$ is somewhat of a normalizing constant as it represents the sum over all edge-weights within the network.

The outcome of the Louvain algorithm is controlled by a resolution parameter, which ultimately affects the number and size of the communites. Larger values lead to fewer but bigger communities and small values produce more sub-groups of smaller size. The optimal choice depends on the specific use case, but in context of the Tweet-Data a value of 0.3 was chosen to enhance smaller partitions.

In figure \ref{fig:graph_communities} some communities are sketched and undoubtedly the detection worked seemingly well. Conclusively, community detection in hashtag graphs are a straight forward methodology to identify topics in Twitter data. However, so far those topics are only existing in the hashtag-graph and still have to be reassigned to the original Tweets. In the upcoming section an approach will be proposed on how to do this reassignments and integrate it into semi-supervised NMF. 

\section{Topic Modelling and Semi-Supervised NMF}

To transfer the identified communities back to the original Tweet data a simple multi-labeling approach is followed, where the community belonging for each hashtag is stored in a look-up table. One can then simply scan through all Tweets and assign the community for each contained hashtag. Hence, each document can end up with zero, one or several labels.

For semi-supervised NMF various models have been developed. We use the topic supervised NMF (TSNMF) implementation of MacMillan et al. \cite{Kelsey.2017} due to its easy-to-use API and the intuitive understanding. Within its core the constraint matrix $L\in \mathbb{R}^{n \times k}$ takes care of the labels 
$$L_{ij} = \left\{
	\begin{array}{ll}
		1  & \mbox{document $d_i$ contains a hashtag of community $j$ } \\
		 & \mbox{or is unlabeled}\\
		0 & \mbox{otherwise}
	\end{array}
\right.$$
Hence the minimization problem changes to
$$ (W,H) = arg ~min \left \| \widetilde{X} - (W\odot  L)H \right \|_F ~ with ~ W, H \geq  0,$$
where $\odot$ represents the Hadamard product.

Loosely speaking, this imposes the factorization not to strive towards the pure reconstruction error anymore, but tries to keep tweets that are assigned to the same community together. For labeled Tweets this is done by eliminating all coefficients of $W$ except the ones for the respective communities during the minimization and thus those components are encouraged to split up the information of the Tweets among them.

\section{ Application on Tweet Data} \label{application}
\subsection{Setup}
The data was collected in the time period between 25th October 2020 to 14th January 2021 with the help of the python package Tweepy \cite{Roesslein.2020} and with a geo-location filter, that covers the United Kingdom. Tweepy connects to the Twitter sample stream, which gives access to a random 1-percent sub-sample of all Tweets published in the given area in real time. To ensure that the Tweets actually contain user generated information that is suitable for topic modelling additional filtering were applied to exclude all Tweets which were either Retweets, written in response to another Tweet or shorter than 160 characters.

For the estimations we initialize the topic model with $k=80$ components and we select the 70 biggest communities for the labeling. The parameter for $k$ was chosen heuristically by trying out different values and $k$'s in the magnitude of 80 in respect to the number of available documents proved to work quite well for both NMF as well as TSNMF.  The number of communities was chosen based on $k$, such that the vast majority of topics are pre-initialized through the labeling, but also such that 10 communities are left uninitialized and can evolve freely. This especially takes  into account that not all topics might be detected through the communities and that some documents can be considered as noise since they are not related to conceptual topics at all. However, it is worth noticing that the labels are merely suggestions to the algorithms, which can and will be overthrown anyways if this leads to a sufficiently lower reconstruction error.

By using the 70 biggest communities, round about  20 \% of all Tweets could be assigned with at least one label. In the original paper MacMillan et al. \cite{Kelsey.2017} suggested that a weighting scheme could be introduced to control the influence of the labeling onto the topic estimates. However, for the sake of simplicity and due to the large number of documents a random down-sampling strategy on unlabeled Tweets was performed to increase the respective ratio of labeled documents up to 50\%. 
 After all still a fairly large collection of round about 310.000 Tweets remained for the further analysis.

\subsection{Evaluation}
The evaluation of topic models is by definition non-trivial since topics are abstract human concepts where no such thing as a ground truth exists. 
Therefore, we believe that the best method to evaluate topics is actually by manual inspection with common sense and by investigating each topic one by one. Furthermore, many of the events that took place in the considered time period are well-known through media, which also yields a good indicator about the topics that are expected to be found.

To determine the effect of the community-based labeling on the outcome, the modelling procedure was done twice with the exact same data and under the exact same setting, but once without any labeling. The comparison reveals, that many topics are indeed being discovered in both cases, but especially the completely unsupervised procedures have some difficulties to find a clear separation between similar topics. The semi-supervised approach on the other hand can make use of the topic information derived by the community detection and leads to more consistent and intuitive results.
This finding is demonstrated in the example in table \ref{tab:comparison} where the topics about the US-election and \textit{Brexit} are characterized by their top 20 words.

\begin{table}[H]
\begin{tabular}{|l|l|}
\hline
                     & \multicolumn{1}{c|}{\textbf{Unsupervised NMF}}                                                                                                                                                                                    \\ 
                     \hline
\textbf{Brexit}      & \begin{tabular}[c]{@{}l@{}}
"uk" 
"brexit" 
"government" 
"eu"
"trump" 
\\
"petition"
"deal"
"trying"
"sign"
"friends"
"says" 
\\ 
"retweet"
"england"
"business" 
"free"
"usa"
"etsy"
\\
"scotland"
"stop"
"times"
\end{tabular}             
\\ 

\hline

\textbf{Mixed}       & \begin{tabular}[c]{@{}l@{}}
"think" 
"trump" 
"say" 
"better"
"thing" 
"bad"
\\
"actually" 
"said" 
"brexit"
"government" 
"wrong" 
\\
"biden" 
"trying"
"twitter" 
"seen" 
"boris" 
"ok"
\\
"believe" 
"imaceleb" 
"maybe"
\end{tabular}            
\\ 

\hline

\textbf{US Election} & \begin{tabular}[c]{@{}l@{}}
"world" 
"trump"
"watching"
"better"
"america"
\\
"country"
"cup"
"kindness"
"biden"
"west"
"man"
\\
"class"
"president"
"place"
"news"
"election2020"
\\
"change" 
"vote" 
"remember" 
"making"
\end{tabular} 
\\ 

\hline

                     & \multicolumn{1}{c|}{\textbf{Semi-Supervised NMF}}                                                                                                                                                                                 \\ 
                     
                     \hline
                     
\textbf{Brexit}      & \begin{tabular}[c]{@{}l@{}}
"uk"
"eu"
"brexit"
"trying"
"deal"
"retweet"
"says"
\\
"government"
"friends" 
"business" 
"post" 
"times"
\\
"tweet" 
"crazy" 
"listening"
"hold"
"johnson"
\\
"coming"
"country" 
"british"
\end{tabular}      \\ 

\hline

\textbf{US Election} & \begin{tabular}[c]{@{}l@{}}
"trump"
"biden"
"election" 
"president" 
\\
"election2020" 
"america" 
"says"
"usa"
"house"
\\
"white"
"media"
"coming" 
"twitter" 
"hold"
\\
"friends" 
"line"
"said"
"results"
"stop"
"big"
\end{tabular}       \\ 

\hline

\end{tabular}%

\caption{Comparison of the topics}
\label{tab:comparison}
\end{table}


It can be seen, that the unsupervised NMF does not only mix up political vocabulary between Brexit and the US-Election, but even completely unrelated noise terms are interfering with the estimates. Here, the Brexit topic is influenced by commercial Tweets regarding the e-commerce platform "Etsy", the mixed topic contains "imaceleb" (TV-show) and the election topic is mixed up with Tweets about the world cup. The community induced semi-supervised topic model on the other hand leads to a clear improvement and both topics are more intuitive to understand.

\section{Conclusion}
Topic modelling in general becomes more challenging when it comes to short and sparse text data, since word-to-word co-occurrences are harder to estimate and additional information is required to derive good results. 
Hashtag graphs in connection with community detection can be used to provide this information as they have been shown to capture a lot of the underlying topic structure of the data. 
For integrating this prior knowledge into the process of topic modelling semi-supervised non-negative matrix factorization can be used as a great interface as it also allows to control the impact of the prior knowledge onto the estimate. 
In an empirical study it was shown that this procedure can help to improve the results to a great extent, as it makes the estimates more robust against noise and helps the algorithm to find a clear separation between topics. This approach ultimately gives back the control of an otherwise unsupervised procedure to the user as both, the granularity of the communities as well as their impact on the topic modelling can be adjusted according to the specific use case.

\bibliographystyle{ACM-Reference-Format}
\bibliography{references}


\begin{thebibliography}{18}


\ifx \showCODEN    \undefined \def \showCODEN     #1{\unskip}     \fi
\ifx \showDOI      \undefined \def \showDOI       #1{#1}\fi
\ifx \showISBNx    \undefined \def \showISBNx     #1{\unskip}     \fi
\ifx \showISBNxiii \undefined \def \showISBNxiii  #1{\unskip}     \fi
\ifx \showISSN     \undefined \def \showISSN      #1{\unskip}     \fi
\ifx \showLCCN     \undefined \def \showLCCN      #1{\unskip}     \fi
\ifx \shownote     \undefined \def \shownote      #1{#1}          \fi
\ifx \showarticletitle \undefined \def \showarticletitle #1{#1}   \fi
\ifx \showURL      \undefined \def \showURL       {\relax}        \fi
\providecommand\bibfield[2]{#2}
\providecommand\bibinfo[2]{#2}
\providecommand\natexlab[1]{#1}
\providecommand\showeprint[2][]{arXiv:#2}

\bibitem[\protect\citeauthoryear{{Aric A. Hagberg}, {Daniel A. Schult}, and
  {Pieter J. Swart}}{{Aric A. Hagberg} et~al\mbox{.}}{2008}]%
        {AricA.Hagberg.2008}
\bibfield{author}{\bibinfo{person}{{Aric A. Hagberg}}, \bibinfo{person}{{Daniel
  A. Schult}}, {and} \bibinfo{person}{{Pieter J. Swart}}.}
  \bibinfo{year}{2008}\natexlab{}.
\newblock \showarticletitle{Exploring Network Structure, Dynamics, and Function
  using NetworkX}. In \bibinfo{booktitle}{\emph{Proceedings of the 7th Python
  in Science Conference}}, \bibfield{editor}{\bibinfo{person}{{Ga{\"e}l
  Varoquaux}}, \bibinfo{person}{{Travis Vaught}}, {and}
  \bibinfo{person}{{Jarrod Millman}}} (Eds.). \bibinfo{address}{Pasadena, CA
  USA}, \bibinfo{pages}{11--15}.
\newblock


\bibitem[\protect\citeauthoryear{Blei, Ng, and Jordan}{Blei
  et~al\mbox{.}}{2003}]%
        {Blei.2003}
\bibfield{author}{\bibinfo{person}{David~M Blei}, \bibinfo{person}{Andrew~Y
  Ng}, {and} \bibinfo{person}{Michael~I Jordan}.}
  \bibinfo{year}{2003}\natexlab{}.
\newblock \showarticletitle{Latent dirichlet allocation}.
\newblock \bibinfo{journal}{\emph{the Journal of machine Learning research}}
  \bibinfo{volume}{3} (\bibinfo{year}{2003}), \bibinfo{pages}{993--1022}.
\newblock


\bibitem[\protect\citeauthoryear{C{\'e}dric and J{\'e}r{\^o}me}{C{\'e}dric and
  J{\'e}r{\^o}me}{2010}]%
        {Cedric.2010}
\bibfield{author}{\bibinfo{person}{F{\'e}votte C{\'e}dric} {and}
  \bibinfo{person}{Idier J{\'e}r{\^o}me}.} \bibinfo{year}{2010}\natexlab{}.
\newblock \showarticletitle{Algorithms for nonnegative matrix factorization
  with the beta-divergence}.
\newblock \bibinfo{journal}{\emph{CoRR}}  \bibinfo{volume}{abs/1010.1763}
  (\bibinfo{year}{2010}).
\newblock


\bibitem[\protect\citeauthoryear{Chen, Zhang, Liu, Ye, and Lin}{Chen
  et~al\mbox{.}}{2019}]%
        {Chen.2019}
\bibfield{author}{\bibinfo{person}{Yong Chen}, \bibinfo{person}{Hui Zhang},
  \bibinfo{person}{Rui Liu}, \bibinfo{person}{Zhiwen Ye}, {and}
  \bibinfo{person}{Jianying Lin}.} \bibinfo{year}{2019}\natexlab{}.
\newblock \showarticletitle{Experimental explorations on short text topic
  mining between LDA and NMF based Schemes}.
\newblock \bibinfo{journal}{\emph{Knowledge-Based Systems}}
  \bibinfo{volume}{163} (\bibinfo{year}{2019}), \bibinfo{pages}{1--13}.
\newblock
\showISSN{0950-7051}
\urldef\tempurl%
\url{https://doi.org/10.1016/j.knosys.2018.08.011}
\showDOI{\tempurl}


\bibitem[\protect\citeauthoryear{Kelsey and James}{Kelsey and James}{2017}]%
        {Kelsey.2017}
\bibfield{author}{\bibinfo{person}{MacMillan Kelsey} {and}
  \bibinfo{person}{D.~Wilson James}.} \bibinfo{year}{2017}\natexlab{}.
\newblock \bibinfo{title}{Topic supervised non-negative matrix factorization}.
\newblock
\newblock


\bibitem[\protect\citeauthoryear{{Mathieu Bastian}, {Sebastien Heymann}, and
  {Mathieu Jacomy}}{{Mathieu Bastian} et~al\mbox{.}}{2009}]%
        {MathieuBastian.2009}
\bibfield{author}{\bibinfo{person}{{Mathieu Bastian}},
  \bibinfo{person}{{Sebastien Heymann}}, {and} \bibinfo{person}{{Mathieu
  Jacomy}}.} \bibinfo{year}{2009}\natexlab{}.
\newblock \bibinfo{title}{Gephi: An Open Source Software for Exploring and
  Manipulating Networks}.
\newblock
\newblock
\urldef\tempurl%
\url{http://www.aaai.org/ocs/index.php/ICWSM/09/paper/view/154}
\showURL{%
\tempurl}


\bibitem[\protect\citeauthoryear{Mazarura and De~Waal}{Mazarura and
  De~Waal}{2016}]%
        {Maz16}
\bibfield{author}{\bibinfo{person}{J. Mazarura} {and} \bibinfo{person}{A.
  De~Waal}.} \bibinfo{year}{2016}\natexlab{}.
\newblock \showarticletitle{A comparison of the performance of latent Dirichlet
  allocation and the Dirichlet multinomial mixture model on short text.}
\newblock \bibinfo{journal}{\emph{In 2016 Pattern Recognition Association of
  South Africa and Robotics and Mechatronics International Conference
  (PRASA-RobMech)}} (\bibinfo{year}{2016}), \bibinfo{pages}{1--6}.
\newblock


\bibitem[\protect\citeauthoryear{Mazarura, De~Waal, and de~Villiers}{Mazarura
  et~al\mbox{.}}{2020}]%
        {Maz20}
\bibfield{author}{\bibinfo{person}{Jocelyn Mazarura}, \bibinfo{person}{Alta
  De~Waal}, {and} \bibinfo{person}{Pieter de Villiers}.}
  \bibinfo{year}{2020}\natexlab{}.
\newblock \showarticletitle{A Gamma-Poisson Mixture Topic Model for Short
  Text}.
\newblock \bibinfo{journal}{\emph{Mathematical Problems in Engineering}}
  \bibinfo{volume}{2020} (\bibinfo{year}{2020}).
\newblock


\bibitem[\protect\citeauthoryear{Mehrotra, Sanner, Buntine, and Xie}{Mehrotra
  et~al\mbox{.}}{2013}]%
        {Mehrotra.07282013}
\bibfield{author}{\bibinfo{person}{Rishabh Mehrotra}, \bibinfo{person}{Scott
  Sanner}, \bibinfo{person}{Wray Buntine}, {and} \bibinfo{person}{Lexing Xie}.}
  \bibinfo{year}{2013}\natexlab{}.
\newblock \showarticletitle{Improving LDA topic models for microblogs via tweet
  pooling and automatic labeling}. In \bibinfo{booktitle}{\emph{Proceedings of
  the 36th international ACM SIGIR conference on Research and development in
  information retrieval}}, \bibfield{editor}{\bibinfo{person}{Gareth~J.F
  Jones}, \bibinfo{person}{P{\'a}raic Sheridan}, \bibinfo{person}{Diane Kelly},
  \bibinfo{person}{Maarten de~Rijke}, {and} \bibinfo{person}{Tetsuya Sakai}}
  (Eds.). \bibinfo{publisher}{ACM}, \bibinfo{address}{New York, NY, USA},
  \bibinfo{pages}{889--892}.
\newblock
\showISBNx{9781450320344}
\urldef\tempurl%
\url{https://doi.org/10.1145/2484028.2484166}
\showDOI{\tempurl}


\bibitem[\protect\citeauthoryear{Pedregosa, Varoquaux, Gramfort, Michel,
  Thirion, Grisel, Blondel, Prettenhofer, Weiss, Dubourg, Vanderplas, Passos,
  Cournapeau, Brucher, Perrot, and Duchesnay}{Pedregosa et~al\mbox{.}}{2011}]%
        {Pedregosa.2011}
\bibfield{author}{\bibinfo{person}{F. Pedregosa}, \bibinfo{person}{G.
  Varoquaux}, \bibinfo{person}{A. Gramfort}, \bibinfo{person}{V. Michel},
  \bibinfo{person}{B. Thirion}, \bibinfo{person}{O. Grisel},
  \bibinfo{person}{M. Blondel}, \bibinfo{person}{P. Prettenhofer},
  \bibinfo{person}{R. Weiss}, \bibinfo{person}{V. Dubourg}, \bibinfo{person}{J.
  Vanderplas}, \bibinfo{person}{A. Passos}, \bibinfo{person}{D. Cournapeau},
  \bibinfo{person}{M. Brucher}, \bibinfo{person}{M. Perrot}, {and}
  \bibinfo{person}{E. Duchesnay}.} \bibinfo{year}{2011}\natexlab{}.
\newblock \showarticletitle{Scikit-learn: Machine Learning in Python}.
\newblock \bibinfo{journal}{\emph{Journal of Machine Learning Research}}
  \bibinfo{volume}{12} (\bibinfo{year}{2011}), \bibinfo{pages}{2825--2830}.
\newblock


\bibitem[\protect\citeauthoryear{Roberts, Stewart, Tingley, Lucas, Leder-Luis,
  Gadarian, Albertson, and Rand}{Roberts et~al\mbox{.}}{2014}]%
        {roberts2014structural}
\bibfield{author}{\bibinfo{person}{Margaret~E Roberts},
  \bibinfo{person}{Brandon~M Stewart}, \bibinfo{person}{Dustin Tingley},
  \bibinfo{person}{Christopher Lucas}, \bibinfo{person}{Jetson Leder-Luis},
  \bibinfo{person}{Shana~Kushner Gadarian}, \bibinfo{person}{Bethany
  Albertson}, {and} \bibinfo{person}{David~G Rand}.}
  \bibinfo{year}{2014}\natexlab{}.
\newblock \showarticletitle{Structural topic models for open-ended survey
  responses}.
\newblock \bibinfo{journal}{\emph{American Journal of Political Science}}
  \bibinfo{volume}{58}, \bibinfo{number}{4} (\bibinfo{year}{2014}),
  \bibinfo{pages}{1064--1082}.
\newblock


\bibitem[\protect\citeauthoryear{Roesslein}{Roesslein}{2020}]%
        {Roesslein.2020}
\bibfield{author}{\bibinfo{person}{Joshua Roesslein}.}
  \bibinfo{year}{2020}\natexlab{}.
\newblock \showarticletitle{Tweepy: Twitter for Python!}
\newblock \bibinfo{journal}{\emph{URL: https://github.com/tweepy/tweepy}}
  (\bibinfo{year}{2020}).
\newblock


\bibitem[\protect\citeauthoryear{Rosen-Zvi, Griffiths, Steyvers, and
  Smyth}{Rosen-Zvi et~al\mbox{.}}{2012}]%
        {rosen2012}
\bibfield{author}{\bibinfo{person}{Michal Rosen-Zvi}, \bibinfo{person}{Thomas
  Griffiths}, \bibinfo{person}{Mark Steyvers}, {and} \bibinfo{person}{Padhraic
  Smyth}.} \bibinfo{year}{2012}\natexlab{}.
\newblock \showarticletitle{The author-topic model for authors and documents}.
\newblock \bibinfo{journal}{\emph{arXiv preprint arXiv:1207.4169}}
  (\bibinfo{year}{2012}).
\newblock


\bibitem[\protect\citeauthoryear{Telesford, Joyce, Hayasaka, Burdette, and
  Laurienti}{Telesford et~al\mbox{.}}{2011}]%
        {telesford2011ubiquity}
\bibfield{author}{\bibinfo{person}{Qawi~K Telesford}, \bibinfo{person}{Karen~E
  Joyce}, \bibinfo{person}{Satoru Hayasaka}, \bibinfo{person}{Jonathan~H
  Burdette}, {and} \bibinfo{person}{Paul~J Laurienti}.}
  \bibinfo{year}{2011}\natexlab{}.
\newblock \showarticletitle{The ubiquity of small-world networks}.
\newblock \bibinfo{journal}{\emph{Brain connectivity}} \bibinfo{volume}{1},
  \bibinfo{number}{5} (\bibinfo{year}{2011}), \bibinfo{pages}{367--375}.
\newblock


\bibitem[\protect\citeauthoryear{{Vincent D Blondel}, {Jean-Loup Guillaume},
  {Renaud Lambiotte}, and {Etienne Lefebvre}}{{Vincent D Blondel}
  et~al\mbox{.}}{2008}]%
        {VincentDBlondel.2008}
\bibfield{author}{\bibinfo{person}{{Vincent D Blondel}},
  \bibinfo{person}{{Jean-Loup Guillaume}}, \bibinfo{person}{{Renaud
  Lambiotte}}, {and} \bibinfo{person}{{Etienne Lefebvre}}.}
  \bibinfo{year}{2008}\natexlab{}.
\newblock \showarticletitle{Fast unfolding of communities in large networks}.
\newblock \bibinfo{journal}{\emph{Journal of Statistical Mechanics: Theory and
  Experiment}} \bibinfo{volume}{2008}, \bibinfo{number}{10}
  (\bibinfo{year}{2008}), \bibinfo{pages}{P10008}.
\newblock
\urldef\tempurl%
\url{https://doi.org/10.1088/1742-5468/2008/10/p10008}
\showDOI{\tempurl}


\bibitem[\protect\citeauthoryear{Wang, Wei, Liu, Zhou, and Zhang}{Wang
  et~al\mbox{.}}{2011}]%
        {wang2011topic}
\bibfield{author}{\bibinfo{person}{Xiaolong Wang}, \bibinfo{person}{Furu Wei},
  \bibinfo{person}{Xiaohua Liu}, \bibinfo{person}{Ming Zhou}, {and}
  \bibinfo{person}{Ming Zhang}.} \bibinfo{year}{2011}\natexlab{}.
\newblock \showarticletitle{Topic sentiment analysis in twitter: a graph-based
  hashtag sentiment classification approach}. In
  \bibinfo{booktitle}{\emph{Proceedings of the 20th ACM international
  conference on Information and knowledge management}}.
  \bibinfo{pages}{1031--1040}.
\newblock


\bibitem[\protect\citeauthoryear{Wang, Liu, Qu, Huang, Chen, and Feng}{Wang
  et~al\mbox{.}}{2014}]%
        {wang2014hashtag}
\bibfield{author}{\bibinfo{person}{Yuan Wang}, \bibinfo{person}{Jie Liu},
  \bibinfo{person}{Jishi Qu}, \bibinfo{person}{Yalou Huang},
  \bibinfo{person}{Jimeng Chen}, {and} \bibinfo{person}{Xia Feng}.}
  \bibinfo{year}{2014}\natexlab{}.
\newblock \showarticletitle{Hashtag graph based topic model for tweet mining}.
  In \bibinfo{booktitle}{\emph{2014 IEEE International Conference on Data
  Mining}}. IEEE, \bibinfo{pages}{1025--1030}.
\newblock


\bibitem[\protect\citeauthoryear{Yin and Wang}{Yin and Wang}{2014}]%
        {Yin14}
\bibfield{author}{\bibinfo{person}{J. Yin} {and} \bibinfo{person}{J. Wang}.}
  \bibinfo{year}{2014}\natexlab{}.
\newblock \showarticletitle{A dirichlet multinomial mixture model-based
  approach for short text clustering.}
\newblock \bibinfo{journal}{\emph{In Proceedings of the 20th ACM SIGKDD
  international conference on Knowledge discovery and data mining}}
  (\bibinfo{year}{2014}), \bibinfo{pages}{233--242}.
\newblock


\end{thebibliography}

\appendix

\end{document}